\documentclass[conference]{IEEEtran}
\IEEEoverridecommandlockouts

\usepackage{cite}
\usepackage{amsmath,amssymb,amsfonts}
\usepackage{algorithmic}
\usepackage{graphicx}
\usepackage{textcomp}
\usepackage{xcolor}
\usepackage{comment}
\usepackage{todonotes}
\usepackage{xspace}
\usepackage{makecell}

\newcommand{\KMELM}{km-scale ELM\xspace}

\def\BibTeX{{\rm B\kern-.05em{\sc i\kern-.025em b}\kern-.08em
    T\kern-.1667em\lower.7ex\hbox{E}\kern-.125emX}}

\begin{document}

\title{Kilometer-Scale E3SM Land Model Simulation over North America
}


\author{\IEEEauthorblockN{Dali Wang, Peter Schwartz, Fengming Yuan, Danial Ricciuto, Peter Thornton, Shih-Chieh Kao, Michele Thornton}
\IEEEauthorblockA{\textit{Environmental Sciences Division},
\textit{Oak Ridge National Laboratory}
Oak Ridge, TN 37831 USA \\
\{wangd,schwartzpd, yuanf, ricciudodm, thorntonpe, kaos, thorntonmm\}@ornl.gov}
\and
\IEEEauthorblockN{Chen Wang, Kathryn Mohror}
\IEEEauthorblockA{\textit{Center for Applied Scientific Computing} \\
\textit{Lawrence Livermore National Laboratory}\\
Livermore, CA USA \\
\{wang116, mohror1\}@llnl.gov}
\and
\IEEEauthorblockN{Qinglei Cao}
\IEEEauthorblockA{\textit{Department of Computer Science} \\
\textit{Saint Louis University}\\
Saint Louis, MO, USA \\
qinglei.cao@slu.edu}
\and
\IEEEauthorblockN{Jayesh Krishnar, Danqing Wu}
\IEEEauthorblockA{\textit{Mathematics and Computer Sciences} \\
\textit{Argonne National Laboratory}\\
Argonne, IL USA \\
\{jayesh, wuda\}@mcs.anl.gov}
}

\maketitle
\thispagestyle{plain}
\pagestyle{plain}
\begin{abstract}
The development of a kilometer-scale E3SM Land Model (km-scale ELM) is an integral part of the E3SM project, which seeks to advance energy-related Earth system science research with state-of-the-art modeling and simulation capabilities on exascale computing systems.
Through the utilization of high-fidelity data products, such as atmospheric forcing and soil properties, the km-scale ELM plays a critical role in accurately modeling geographical characteristics and extreme weather occurrences. The model is vital for enhancing our comprehension and prediction of climate patterns, as well as their effects on ecosystems and human activities. 

This study showcases the first set of full-capability, km-scale ELM simulations over various computational domains, including simulations encompassing 21.6 million land gridcells, reflecting approximately 21.5 million square kilometers of North America at a 1 km $\times$ 1 km resolution. 
We present the largest km-scale ELM simulation using up to 100,800 CPU cores across 2,400 nodes. This continental-scale simulation is 300 times larger than any previous studies, and the computational resources used are about 400 times larger than those used in prior efforts. Both strong and weak scaling tests have been conducted, revealing exceptional performance efficiency and resource utilization. 

The km-scale ELM uses the common E3SM modeling infrastructure and a general data toolkit known as KiloCraft. Consequently, it can be readily adapted for both fully-coupled E3SM simulations and data-driven simulations over specific areas, ranging from a single gridcell to the entire North America. 
 

\end{abstract}



\section{Introduction}
Earth System Models (ESMs) play a vital role in advancing our understanding of the Earth's climate system and its response to natural and human-induced changes. These models incorporate various components such as the atmosphere, oceans, and land surface, along with their dynamic interactions, to provide valuable insights into past, present, and future climate conditions. ESMs contribute to climate research, policy-making, and public awareness, enabling us to make informed decisions and take proactive measures to address the challenges posed by climate change. The development and application of ESMs require significant computational resources and expertise. 

The Energy Exascale Earth System Model (E3SM), funded by the US Department of Energy (DOE), was developed to align with the nation's scientific and the DOE's mission objectives. E3SM distinguishes itself from many other ESMs by emphasizing energy-related research. This includes investigating the effects of climate change on energy systems, such as water availability, severe weather events, and sea level rise~\cite{golaz2019doe}. E3SM's key advantages include its high-resolution capability, allowing for better representation of smaller-scale processes like regional weather patterns and ocean currents. The model is optimized for advanced exascale computers, enabling complex simulations that would be computationally prohibitive on older systems, leading to more detailed and accurate predictions~\cite{caldwell2019}. 

 E3SM Land Model (ELM)\cite{burrows2020doe} provides detailed representations of vegetation, soil moisture, surface temperature, and land-use changes. ELM interacts with other E3SM components, such as the atmosphere and oceans, to gain insights into how these elements influence each other and the overall Earth system. Key processes modeled by ELM include energy and water exchange, carbon cycling, vegetation dynamics, and human activities like agriculture and deforestation. ELM is crucial for understanding surface conditions that influence atmospheric circulation and climate patterns, making them valuable tools for studying climate change impacts, land-use change scenarios, and carbon sequestration potential.

\subsection{Related work}

The majority of ESM simulations are conducted on coarse grids larger than 0.5 $\times$ 0.5 degrees~\cite{soares2024high}. The Energy Exascale Earth System Model version 1 (E3SMv1) simulates at two different horizontal resolutions: a standard resolution with approximately 100 km (around 1 degree) grid spacing for land and atmosphere, and a high-resolution with 25 km (0.25 degree) grid spacing for land and atmosphere. The high-resolution version allows for a detailed representation of ocean eddies and atmospheric storms, which are vital for understanding planetary circulation and their impacts on human activity~\cite{caldwell2019}. E3SM is currently developing kilometer-scale models that fully utilize Exascale computers to provide detailed representations of environmental features. The development of kilometer(km)-scale ELM model is being carried out in conjunction with other E3SM components aimed at exascale computing, such as the kilometer(km)-scale Atmosphere and Ocean models~\cite{Donahue2024,omega}

The ELM software comprises half a million lines of code, featuring highly specialized data types containing around 2,000 globally accessible, multidimensional arrays, and over 1,000 subroutines~\cite{xu2017web,zheng2019xscan}. In recent years, there have been advancements in creating an ultrahigh-resolution ELM (uELM) simulation that leverages exascale computing for accurate land simulations on continental and global scales. For instance, a new computational model and framework for uELM simulation on hybrid architectures of exascale computers have been introduced~\cite{wang2022towards,YUAN2023102145}. Various strategies for porting uELM have been established, alongside the development of a functional unit test (FUT) framework~\cite{schwartz2022spel,schwartz2022developing}. These uELM simulations were carried out on a small domain with fewer than 10,000 grid cells, utilizing small-scale HPC clusters with only a few nodes (less than 5). An ELM simulation at a 1 km by 1 km resolution over a small domain, containing fewer than 10,000 grid cells, were reported in 2022. However, that study employed a simplified ELM model (with a satellite phenology submodel and a reduced number of subgrid components) to understand the sub-grid topographic effects on land-atmosphere interactions over mountainous areas~\cite{hao2022impacts}.

With the availability of high-fidelity data products, such as the forcing and soil properties~\cite{thornton2014daymet,thornton2021gridded,han2023global}, km-scale ELM can deliver high spatial resolution beneficial for accurately simulating geographical features and extreme weather events. However, the inherent computational challenges pose a significant barrier to their implementation. Specifically, the increased grid density demands advanced computing resources and storage capacities to handle the vast data volumes.

\subsection{Km-scale ELM simulation}
To effectively tackle the computational challenges associated with km-scale ELM simulations, we have developed methods that optimize the model configuration and enhance data handling and processing capabilities. The study delivered an unprecedented ELM simulation capability, with its contributions described as follows:
\begin{enumerate}
\item The study conducted the largest km-scale ELM simulation to date. This simulation, which operated at full ELM capacity using a common E3SM modeling framework, covered a continental scale and was 300 times larger than any previous study reported.

\item The study presented the largest km-scale ELM simulation using up to 100,800 CPU cores across 2,400 nodes. The study archived outstanding scaling performance by utilizing computational resources that were about 400 times larger than those used in prior efforts.  

\item This study developed a data toolkit to support ELM simulation at various scales, ranging from a single gridcell to the entire North America domain of 21.5 million gridcells. This study also created the first 1 km by 1 km dataset (including domain, surface properties, and forcing) for ELM simulations across North America.  
\end{enumerate}

\section{Method}

 The following four sections outline our method: model configuration, input data generation, numerical experiments to assess the ELM simulation readiness, and results validation and performance profiling. 

\subsection{Data-driven, km-scale ELM configuration and simulation}

The Common Infrastructure for Modeling the Earth (CIME) is a crucial framework in the E3SM project, which supports the development, simulation, and testing of Earth system models. CIME offers a standardized toolkit, procedures, and approaches to improve and simplify model development for different elements. Within our research, we employed CIME scripts for configuring, constructing, and submitting batch jobs for the km-scale ELM setup. Presently, the km-scale ELM consists of an active land model, a data atmosphere model, and a coupler component integrated through CIME. 
\vspace{2mm}
\subsubsection{Component configuration and dataflow}
\begin{figure}[b]
\centerline{\includegraphics[width=\linewidth]{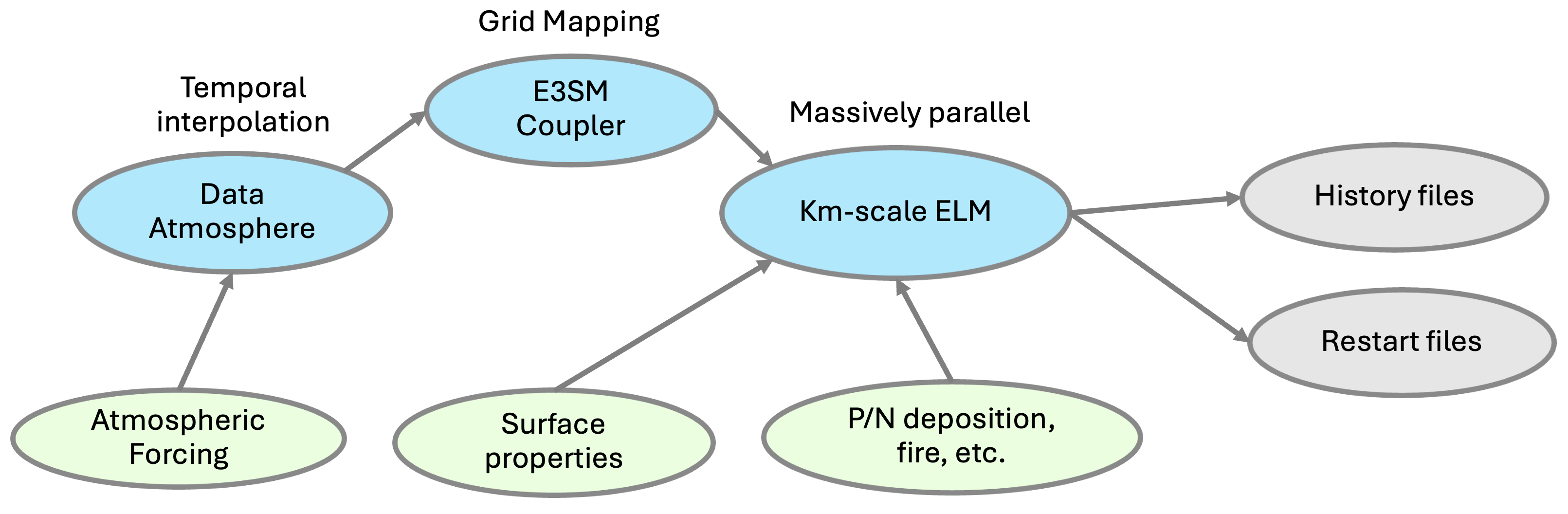}}
\caption{Km-scale ELM data flow and simulation configuration, which requires three groups of datastreams: atmospheric forcing, land surface properties, and others (e.g., nitrogen and phosphorus (N/P) deposition).} 
\label{fig:km-ELM-data-flow}
\end{figure}

The experiment employs a designated component setting known as I1850uELMCNPRDCTCBC. Within this setting, the active E3SM land simulation is driven by predefined atmospheric forcing to attain steady states for transient ELM simulations since the Industrial Revolution (e.g., 1850). User-defined forcing data are retrieved from the disk through a data atmospheric model and data exchange is enabled through a coupler, as illustrated in Fig~\ref{fig:km-ELM-data-flow}. The land model is configured to simulate biophysical and biochemical processes that encompass energy, water, and carbon, nitrogen, and phosphorus (CNP) dynamics under resource demand (RD). Additionally, the experiment setting includes a soil organic matter decomposition model (converging trophic cascade (CTC)) and a black carbon (BC) model. 

The experiments utilize custom grids specified in the domain files, such as a 1 km x 1 km resolution grid covering North America. A new setup was created for the data atmospheric model to retrieve user-defined atmospheric forcing data from the disk. To improve simulation efficiency and avoid spatial grid conversion and interpolation, both the atmosphere and land models share the same domain file.

In situations where the atmospheric forcing data is not available at the same time intervals as required by the land model, the atmosphere model temporally interpolates the data to ensure synchronization during simulation. 

\vspace{2mm}
\subsubsection{Computational domain, domain decomposition, and parallel execution}

The ELM computational domain employs a Lambert conformal conic projection (LCC) taken from the Daymet data product\cite{thornton2021gridded}. This conic map projection, centered at a latitude of 42.5, ensures shape precision, offering an effective portrayal of North America and supporting accurate mapping in meters. The ELM domain also utilizes the WGS84, a global geodetic reference system, to facilitate data exchange with other Earth components. 

The km-scale ELM uses MPI for parallel processing, assigning a fixed number of land gridcells among MPI processes using static domain decomposition. Three partitioning schemes, namely round-robin, block round-robin, and block partition, are implemented. Each MPI process retains a complete copy of the ELM domain information, including all the subdomain dimensions and sizes. However, each MPI process only allocates memory for conducting ELM over its designated subdomain. Within each MPI process, the ELM model acquires atmospheric forcing data from the atmospheric model, utilizes surface characteristics and land-use datasets to initialize land gridcells within its own subdomains, and then conducts parallel simulations across these gridcells independently. Depending on the simulation domain size and hardware setup, km-scale ELM can run on CPUs or GPUs utilizing OpenACC~\cite{schwartz2022spel}. This study exclusively leverages CPU resources, with GPU utilization planned for future work. 

\vspace{2mm}
\subsubsection{I/O and data management}

The ELM provides the capability to generate time-averaged history outputs and restart files at user-defined intervals, which can range from timestep-based, hourly, daily, monthly, annually, or no output at all. 
ELM also allows users to customize output variables beyond the default output variables. The simulation outcomes are saved as historical files (Fig.~\ref{fig:km-ELM-data-flow}). They can be later processed and analyzed by the uELM data management and analysis component, developed based on VISIT~\cite{Childs2012visit} and QGIS~\cite{QGIS_software}. Additionally, a restart function is activated to save files periodically for the active Earth system components including the data atmosphere model, the coupler, and the land model. These restart files can be used to restart the simulation from a specific simulated time point. This is especially useful for long-term simulation or in case of system failures. 

The size of history files depends on several factors: history output frequency, the number of gridcells, top gridcell level variable quantities, and their datatypes. In contrast, the size of restart files is directly proportional to the number of gridcells and their subgrid elements, as well as variable quantities and datatypes at all gridcell levels.

The data produced by the \KMELM models are not decomposed in memory across the compute processes in an ``I/O friendly'' way.
To improve I/O performance, \KMELM uses a specialized parallel I/O library, namely SCORPIO~\cite{scropio2020, Hartnett2021PIO}, to perform data accesses. SCORPIO rearranges the data distributed across multiple processes before passing it to the backend I/O libraries like PnetCDF, NetCDF, and ADIOS to write the data to the file system.
These backend I/O libraries may further optimize I/O by aggregating operations from various MPI processes into smaller groups known as aggregators, which directly interact with the underlay parallel file system, such as Lustre~\cite{sun:2007:lustre} and GPFS~\cite{schmuck:2002:gpfs}.

\subsection{Km-scale ELM input data generation}

Drawing from earlier research in large-scale data processing~\cite{wang2023data}, we have devised a comprehensive data toolkit named KiloCraft. This toolkit is designed to facilitate the generation of suitable input datasets, such as forcing, surface properties, and land use, for kilometer-scale ELM simulations. 

KiloCraft comprises essential functionalities to aid in domain, forcing, and surface data generation at a resolution specified by the user within a defined computational domain, such as North America in this study. Additionally, it offers features for data subsetting in particular areas of interest, duplicate input datasets, sanity checks, cross-data product validation, and visualization. 
\vspace{2mm}
\subsubsection{ELM domain over North America}

We extract the domain information from a Daymet product, such as grid dimensions, variable values, and coordinate information. We calculate gridcell IDs, create land gridcell masks, and determine the area and fraction of land for each gridcell. We also convert coordinates between the LCC projection system and WGS84 coordinate systems and generate 1D and 2D ELM domains over North America (using NetCDF). These steps are essential for setting up the computational domain and preparing the meteorological data for subsequent simulations. 

This ELM domain offers a distinguishable 1 km by 1 km resolution across North America, comprising more than 21.5 million square kilometers of land gridcells. (In ELM simulations, lakes are also considered as land gridcells). 
\vspace{2mm}
\subsubsection{ELM forcing data}

\begin{figure*}[thbp]
\centerline{\includegraphics[width=0.9\linewidth]{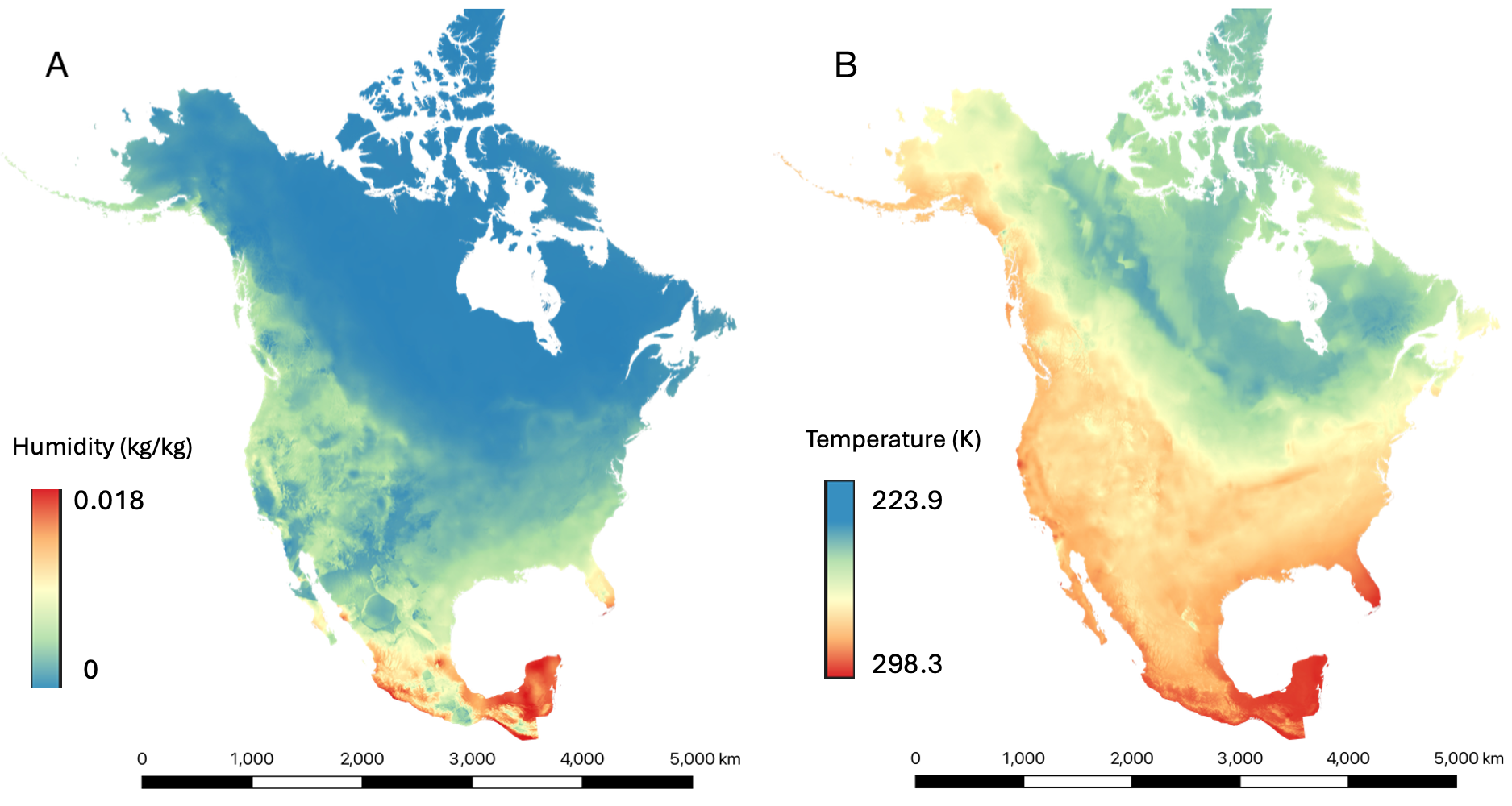}}
\centerline{\includegraphics[width=0.9\linewidth]{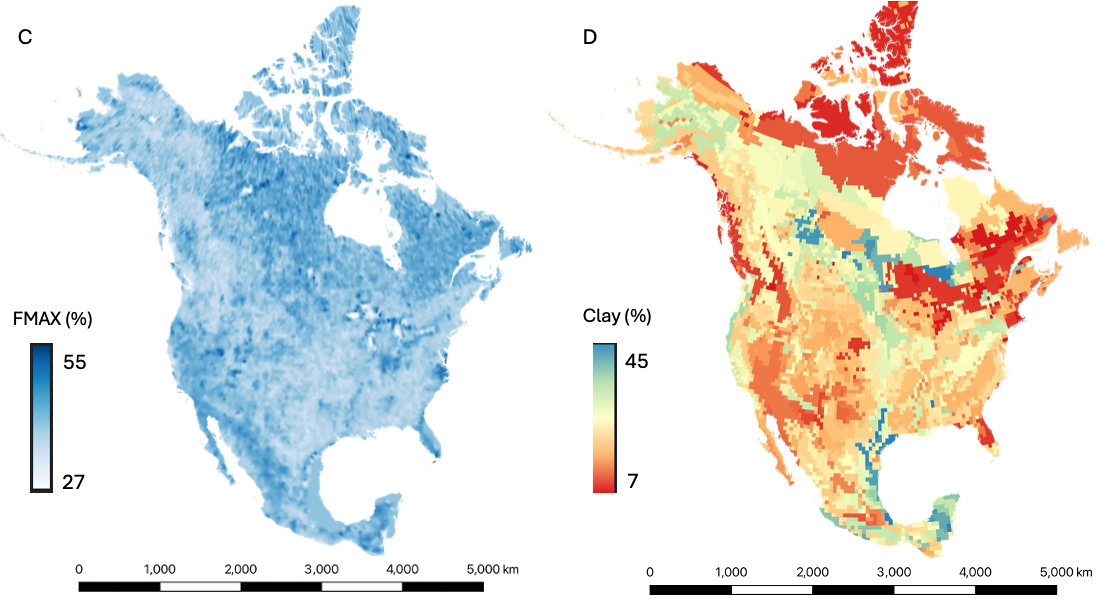}}
\caption{Samples of 1 km by 1 km input data across North America. Top left: humidity (01/01/2014); top right: temperature (01/01/2014); bottom left maximum fractional
saturated area (percentage); bottom right: clay (percentage) in soil.}
\label{fig_2}
\end{figure*}

Our experiments utilize meteorological forcing data from Daymet, which offers gridded estimates of daily weather parameters in North America from 1980 to the present at a 1 km spatial resolution~\cite{thornton2021gridded,kao2024gridded}. Initially, the Daymet data was temporally downscaled to 3-hourly intervals to align with the requirements for ELM simulations. This downsizing process incorporated sub-daily temporal information from the well-established GSWP3 meteorological reanalysis dataset, ensuring consistency in each sub-daily time step and maintaining the overall daily Daymet value. Forcing samples are illustrated in the Fig~\ref{fig_2} upper panels.

Furthermore, KiloCraft is employed to produce one-dimensional meteorological forcing data formatted specifically for ELM. This technique involves creating a Lambert conformal conic projection for geospatial transformations and temporal data processing. An essential step in this process is the development of a land mask to exclude non-land data. Subsequently, the 2D forcing data is reshaped and extracted for valid land cells, eventually stored in a one-dimensional format along with metadata. When feasible, parallel processing is implemented for enhanced efficiency. 

\vspace{2mm}
\subsubsection{ELM surface properties}

KiloCraft is also employed to produce 1 km by 1 km resolution surface property data for ELM simulations. Users can choose interpolation methods (such as nearest neighbor, linear, spline) for individual variables to generate high-resolution data from coarse-resolution gridded data or observational data points collection. Additionally, KiloCraft enables the integration of state-of-the-art data products directly into the surfaces. In our current experimental scenario, we utilize a standard resolution (0.5 degrees x 0.5 degrees) global surface properties dataset in the WSG84 system to produce refined data products within the Daymet domain, represented in both 1D and 2D formats.

Km-scale ELM also requires additional small input files, such as parameter files for ELM/CNP, and snowpack treatment. Given that these parameters do not exhibit a strong correlation with model resolution, we utilize the parameter files from standard ELM input data repositories.

The North America domain has dimensions of 7,814 x 8,075, with 21.6 million gridcells treated as land gridcells. Each gridcell is structured hierarchically, encompassing various subgrid components including topographic units, land units (up to 5 types), soil columns (up to 2 types with 15 layers each), plant function types (up to 17 vegetation types), and cohorts. These components collectively represent the diverse characteristics of North America's surface and subsurface. Sample were illustrated in the Fig.~\ref{fig_2} bottom panels.


\subsection{ELM experiments to evaluate massive simulation readiness}

Kilometer-scale ELM provides high spatial resolutions for capturing intricate interactions in the atmosphere and landscape surface, allowing for the representation of processes often oversimplified in coarse-resolution models. However, apart from issues related to parallel computing and simulations, these models present additional challenges. Particularly, challenges associated with parametrization arise due to the need to represent phenomena such as land-atmosphere interactions and nutrient cycles. Data assimilation and validation necessitate suitable observational datasets and techniques for effectively integrating vast quantities of output data. Understanding interactions between models becomes increasingly intricate at finer scales, leading to challenges in grasping feedback mechanisms. The elaboration of physical and biological characteristics demands a profound understanding of parameter values and interactions, often requiring further investigation. The accurate specification of boundary and initial conditions becomes increasingly crucial, requiring high-quality data at all scales.

Although efforts are underway to tackle these challenges, our study focuses on a specific study domain that has demonstrated success in a previous project, namely the NGEE Arctic project~\cite{ngee-arctic}. Through duplicating this domain, we aim to assess the scalability and input/output challenges of the km-scale ELM simulation framework.

Initially, we established a collection of land gridcells in the Alaska Seward Peninsula (AKSP) region, totaling 72,083 gridcells. Subsequently, we utilize KiloCraft to produce input datasets, including domain, forcing, and surface properties for the AKSP region. We conducted the AKSP experiments on the CADES Baseline system at Oak Ridge National Laboratory (ORNL). The open-access portion of Baseline includes 180 nodes, each with 128 cores and 2X AMD 7713 processors. The AKSP experiment acts as the baseline simulation for our model's development and evaluation. 

\vspace{2mm}
\subsubsection{AKSP baseline simulation}
\begin{figure*}[htbp]
\centerline{\includegraphics[width=0.78\linewidth]{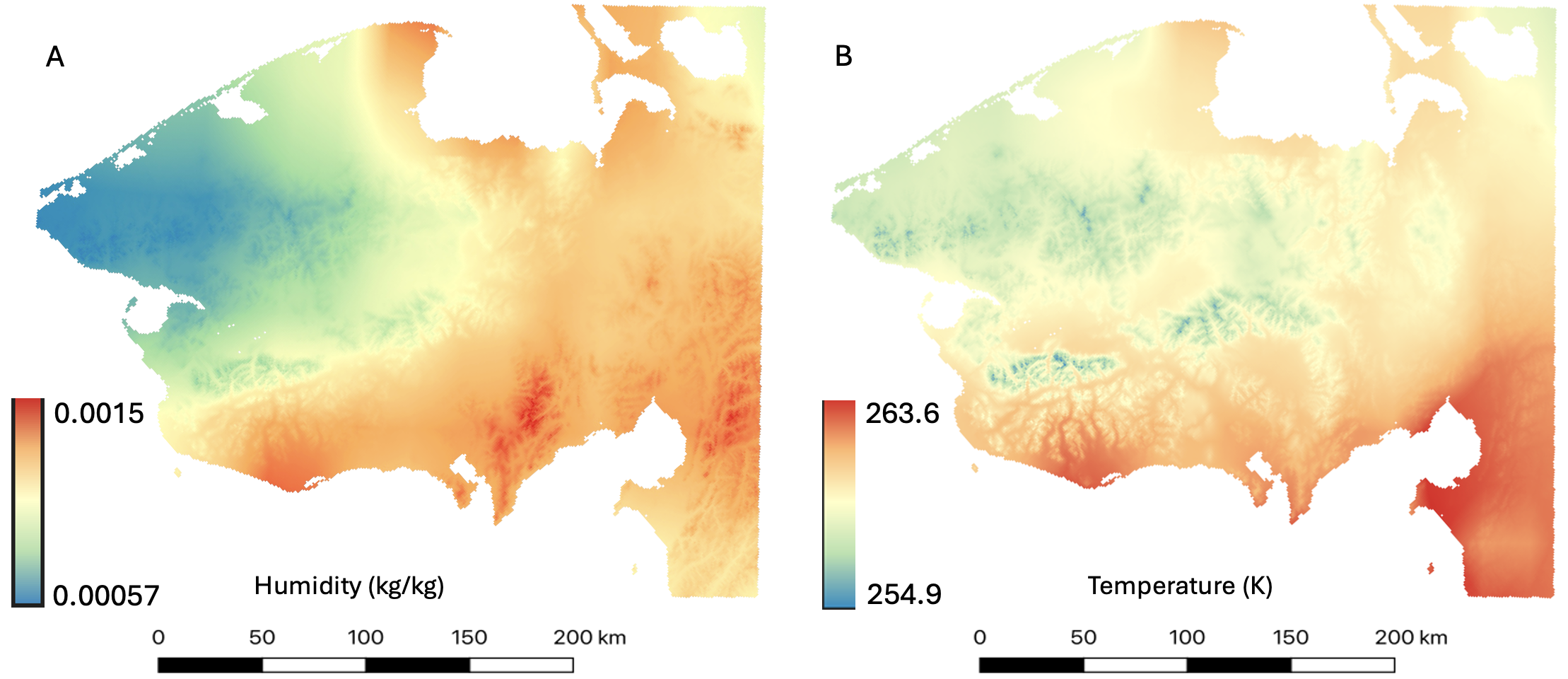}}
\centerline{\includegraphics[width=0.78\linewidth]{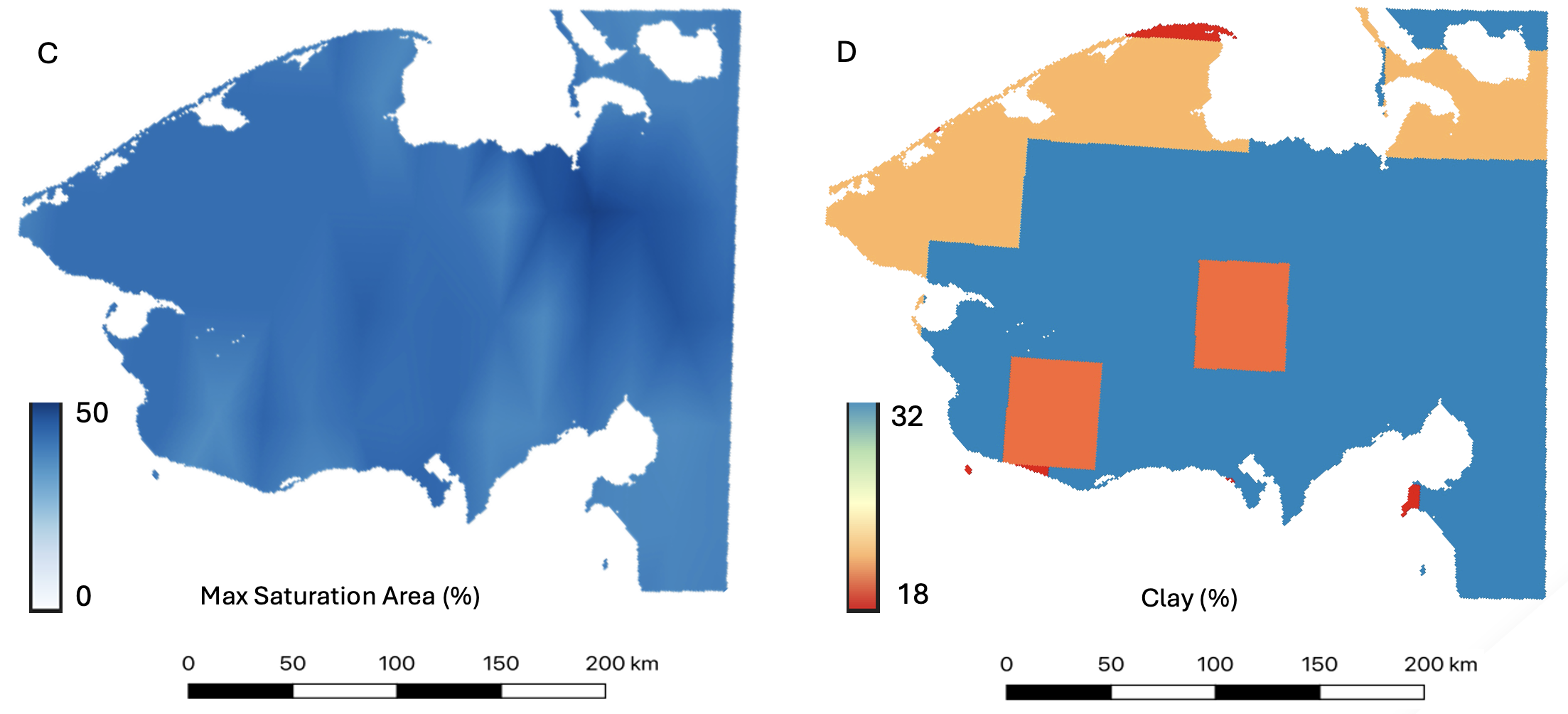}}
\caption{The domain and input dataset of the AKSP reference simulation. Top left: humidity (01/01/2014); top right: temperature (01/01/2014); bottom left: maximum fractional saturated area (percentage); bottom right: clay (percentage) in the soil.}
\label{fig_3}
\end{figure*}

The Seward Peninsula, situated on the western coast of Alaska, is a vast landmass spanning approximately 330 km in length and 145–225 km in width, enveloping a total land area of 72,083 square kilometers. Table~\ref{tab1} lists input data files for the AKSP simulation. For illustration purposes, the domain and sample input variables for the AKSP reference simulation are also presented in Fig.~\ref{fig_3}. The forcing data used in the AKSP experiment is a subset from the North America dataset, which originated from the Daymet data product at the 1 km by 1 km spatial resolution. Meanwhile, the surface properties data was interpolated (with the nearest neighbor method) from a lower resolution dataset at 0.5 degrees by 0.5 degrees. Consequently, variables, such as the clay soil percentage, for the AKSP region exhibit a large box-like pattern at an approximate resolution of 50 km by 50 km. Efforts are underway to integrate and produce a 1 km by 1 km surface properties dataset. Since this study aims to assess the computational capacity of the kilometer-scale ELM, therefore, we used the original data source from previous research. Prior research utilized a simulation framework tailored for point-wise ELM simulation in small areas, covering the AKSP domain with tiles~\cite{YUAN2023102145}. It is noteworthy that the km-scale ELM reference simulation performed approximately four times faster than the simulations used in the cited study.

Km-scale ELM experiments are configured to produce default ELM output history files (elm.h0), encompassing all essential variables over the land gridcells. To maintain experiment consistency, we establish a fixed hourly timestep for a five-day simulation. Subsequently, the output is configured as temporal averaging for the entire simulation period, averaging in a single timestep at the simulation's completion. This setting represents the default ELM model output situation, in which at the end of each simulation month, the model saves temporally (monthly) averaged results as single timestep variables into a NetCDF file. Additionally, the restart interval is set as the simulation period. Consequently, at the simulation's completion, ELM generates restart files for all components, namely, the coupler (cpl.r), the land (elm.r), and the data atmosphere model.
The data atmosphere restart file is a small file with atmospheric data streams. Additionally, ELM will save a small auxiliary land restart history (elm.rh0) file, which is required for the continuation of a simulation. This configuration represents the most demanding I/O workload scenario in ELM simulations. Table~\ref{tab2} lists information on these large output files from the AKSP experiment.

\begin{table}[t]
\caption{Input data files for the AKSP reference simulation}
\begin{center}
\begin{tabular}{|c|c|c|c|c|}
\hline
& \# Variables & File Size & \makecell{Largest\\Dimension} & \makecell{Size of\\Largest Variables} \\
\hline
Domain & 15 & 0.09 GB & \makecell{72,083\\(gridcell)} & \makecell{$4 \times 72,083$\\(double)} \\
\hline
\makecell{Surface\\data} & 89 & 1.63 GB & \makecell{72,083\\(gridcell)} & \makecell{$12 \times 17 \times 72,083$\\(double)} \\
\hline
\makecell{Forcing\\(annual)} & 7 & 6.01 GB & \makecell{72,083\\(gridcell)} & \makecell{$248 \times 72,083$\\ (double)} \\
\hline
\end{tabular}
\label{tab1}
\end{center}
{\footnotesize Double means double precision float-point.}
\end{table}



\begin{table}[b]
\caption{Large output data files from the AKSP reference simulation}
\begin{center}
\begin{tabular}{|c|c|c|c|c|}
\hline
& \# Variables & File Size & \makecell{Largest \\Dimension} & \makecell{Size of \\Largest Variables} \\
\hline
elm.h0 & 553 & 0.48 GB & \makecell{233,1447\\(pft)} &\makecell{$17 \times 72,083$\\(float)} \\
\hline
cpl.r & 92 & 0.05 GB & \makecell{72,083\\(gridcell)} & \makecell{$1 \times 72,083$\\(double)} \\
\hline
elm.r & 466 & 14.18 GB & \makecell{233,1447\\(pft)} & \makecell{$2 \times 2,331,447$\\(double)} \\
\hline
\end{tabular}
\label{tab2}
\end{center}
{\footnotesize Double stands for double precision float-point, and float means single-precision floating-point}
\end{table}

\vspace{2mm}
\subsubsection{Domain duplication for simulation readiness evaluation} 

The experiments are presently configured as a spin-up simulation that includes ecosystem dynamics to reach stable-state values for subsequent transient simulations. This study highlights the ability for parallel simulation capability, km-scale modeling configuration, and the computational performance (computing and I/O). 

It is worth mentioning that the ELM simulation is conducted independently at all individual land gridcells. The simulation is spatially implicit within each cell, with no lateral transfer among gridcells. Therefore we replicate the reference AKSP cases to create numerical experiments of various sizes. These experiments are used to evaluate the computational performance of the km-scale ELM on large-scale computers.

\begin{table}[htb]
\caption{The size of input files of the experiments (unit: GB)}
\begin{center}
\begin{tabular}{|c|c|c|c|c|}
\hline
& \makecell{AKSP} & \makecell{AKSP\\x10} & \makecell{AKSP\\x10x10} & \makecell{AKSP\\x10x10x3} \\
\hline
Domain & 0.01 & 0.1 & 0.9 & 2.8 \\
\hline
Surface data & 1.6 & 7.2 & 63.4 & 188.2 \\
\hline
Forcing (annul) & 7.1 & 50.3 & 497.7 & 1477.4 \\
\hline
\end{tabular}
\label{tab3}
\end{center}
\end{table}

In our study, we replicated AKSP cases to construct ELM experiments of varying sizes, listed in Table~\ref{tab3}. The largest experiment conducted (ASKPx10x10x3) encompasses a domain comprising 21.6 million (21,624,900) land gridcells, matching the total number of land gridcells in North America, as represented within the Daymet domain. Furthermore, the AKSPx10x10x3 domain comprises the following subgrid cell components: 21.6 million (21,624,900) topographic units, 93.9 million (93,936,900) land units, 353.4 million (353,435,700) soil columns, and 699.4 million (699,434,100) plant function types.

\vspace{2mm}
\subsubsection{Scalability test on the Summit supercomputer} 

This study employed Summit, a supercomputer located at ORNL. Summit comprises 4,608 nodes, each containing two 22-core IBM POWER9 CPUs and six Nvidia Tesla GPUs, providing more than 600 GB of coherent memory. Two cores out of the 44 in each node are dedicated to system operations, leaving 42 cores per node for our experiments. The experiments were developed using ncvhp/23.9, NetCDF (C/Fortran), and PnetCDF libraries. 

In practice, ELM simulations can run for extensive periods, sometimes up to hundreds of years. The default output interval for ELM is typically monthly history files, with restart files saved every 10-20 years. In our scalability tests, we configured the ELM simulation to run over a brief 5-day period with an hourly timestep. This ensured that the land component did not overshadow other components. The outcomes of these experiments included one history file and one restart file. These short-term simulations provided essentail information to assess km-ELM computational capabilities, scalability using limited computation allocations.

In this study, we conducted two distinct scalability tests using the reference simulation case, AKSP, which encompasses a total of 72,083 land gridcells. The first test was designed to assess strong scalability, where we employed the configuration AKSP x 10 x 10 x 3 across varying node counts, specifically using 300, 600, 1200, and 2400 nodes, with each node allocated 42 MPI tasks. This configuration allowed us to explore how well the computational performance improves as we increase the number of processing units while keeping the problem size constant. The strong scalability results illustrate the efficiency of our parallel implementation in dealing with larger clusters, thereby indicating how well the workload can be distributed across multiple nodes without significant overhead.

The second test focused on weak scalability, examining how the system performs when the workload is proportionally increased with the addition of more nodes. The configurations included AKSP for 1 node, AKSP x 10 for 10 nodes, AKSP x 10 x 10 for 100 nodes, and AKSP x 10 x 10 x 3 for 300 nodes. This approach emphasizes the ability of the system to maintain performance as the problem size scales with the number of computational resources. By evaluating the weak scalability, we aim to determine the limits of our simulation's capacity to efficiently handle larger-scale scenarios, providing insights into the robustness of our method for extensive environmental simulations.

\subsection{Result validation and performance profiling}

\subsubsection{ELM outputs: history and restart files}

The km-scale ELM simulations produce a single history output file as well as several restart files. The comparison of data within each file is conducted to verify and validate the outputs using duplicated input data. By comparing the output of the AKSP case simulation on Summit with that of the reference AKSP simulation on the baseline system, further evaluation can be achieved. As we utilized two machines, with most variables in double precision within the history and restart files, a bit-by-bit comparison was performed accounting for machine precision to ensure equivalency at the machine level. Technically, to obtain robust performance, the km-scale ELM adopts the PnetCDF library and uses the NETCDF3\_64BIT format.

\vspace{2mm}
\subsubsection{Timing and performance profiling} 

The General Purpose Timing Library (GPTL) integrated into the CIME framework plays a crucial role in our kilometer-scale ELM experiments by collecting detailed model timing data. GPTL timers have been placed within both the CIME driver and model code to capture detailed timings for thorough performance analyses. Three main types of timing data are collected during each model run: 1) comprehensive run metrics such as total run time, cost, and throughput; 2) detailed statistics for individual model timers; and 3) daily runtime tracking information to pinpoint any variations.

By harnessing GPTL, our experiments can guarantee precise timing evaluations, which are paramount for optimizing model performance and scrutinizing resource utilization in our kilometer-scale ELM simulations. We also use the SCROPIO capability to trace and profile I/O performance, including the read/write data size, time, and throughput.

We utilize the Performance Analytics for Computational Experiments (PACE), a framework designed to summarize performance data gathered from E3SM experiments. This framework enables us to extract valuable insights and present them via an interactive web portal. Users can access the PACE web portal at https://pace.ornl.gov/ to explore and analyze the collected performance data.

\section{Result and discussion}

\subsection{Simulation output}

\begin{figure*}[hbtp]
\centerline{\includegraphics[width=0.78\linewidth]{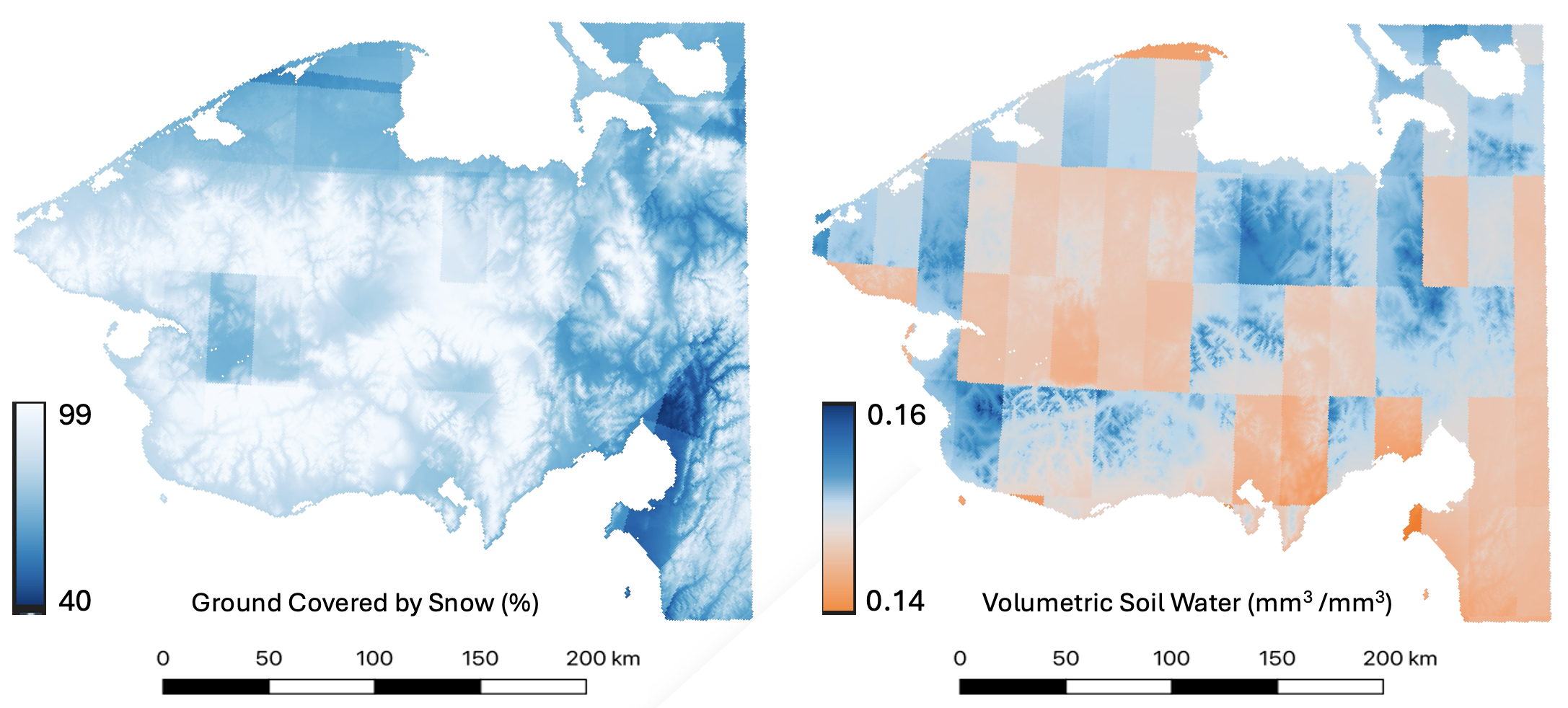}}
\centerline{\includegraphics[width=0.78\linewidth]{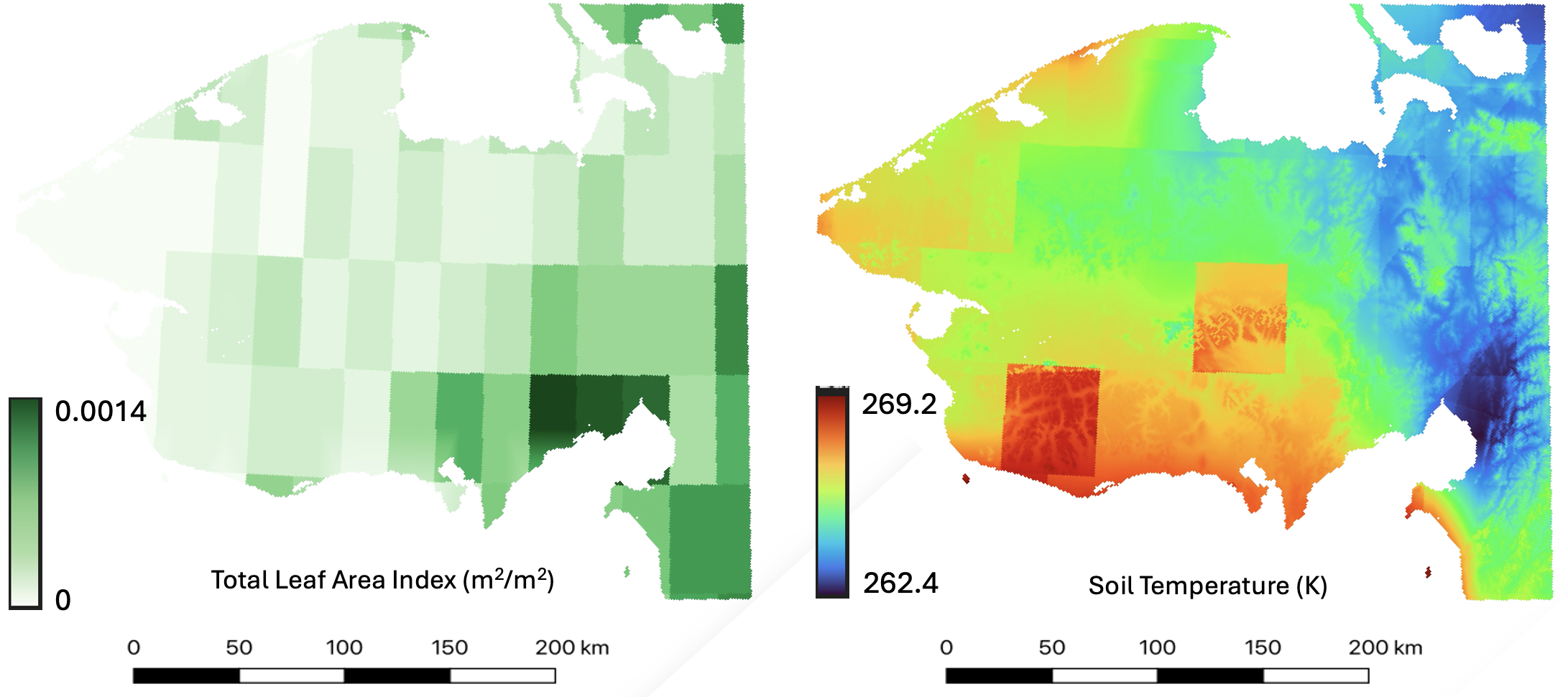}}
\caption{Visualization of output variables from the experiment. Top left: ground coverage by snow; top right: water in soil; bottom left: total leaf area index; bottom right: soil temperature.}
\label{fig_4}
\end{figure*}

Our experiments output one history file and several restart files at the end of each simulation. We have developed a Python-based software utility to facilitate the comparison of variables contained within the AKSP case executed on the Summit supercomputer and the reference results obtained from the baseline system. For each variable identified in both files, the utility performs an element-wise numerical comparison to detect any discrepancies in their data values. Additionally, it reports any variable present in one file but absent in the other. Furthermore, for duplicate AKSP cases, the utility is employed to ascertain that the result precisely replicates the given number of copies of the reference simulation. Fig.~\ref{fig_4} displays four output variables extracted from history files from the AKSP case. These are early simulation results from the spin-up phase, hence the total leaf area index values are very low. Box patterns are observed in the results, arising the discrepancies in the input surface properties dataset. Nonetheless, these outcomes are identical to the reference simulation results from the baseline system.

Table~\ref{tab4} listed the sizes of the history and restart files generated from various experiments. The size of the files is apportioned to the number of gridcells in each experiment, for example, the elm.h0 history file from the AKSPx10x10x3 case is approximately 280 times larger than the h0 file from the AKSP case. The ELM restart file from the largest experiment reaches nearly 4.3 terabytes. 

\begin{table}[b]
\caption{The size of output files from various experiments (unit: GB)}
\begin{center}
\begin{tabular}{|c|c|c|c|c|}
\hline
& AKSP & AKSPx10 & AKSPx10x10 & AKSPx10x10x3 \\
\hline
elm.h0 & 0.48 & 4.81 & 47.91 & 134.18 \\
\hline
elm.r & 14.18 & 141.63 & 1,416.19 & 4,248.54 \\
\hline
cpl.r & 0.05 & 0.49 & 4.75 & 14.20 \\
\hline
\end{tabular}
\label{tab4}
\end{center}
\end{table}

\subsection{Performance evaluation}

In this study, we investigate the two scalability tests with the duplicated reference simulation case, AKSP, which consists of 72,083 land gridcells. The first test evaluated strong scalability by using the configuration AKSP x 10 x 10 x 3 across various node counts. This allowed us to assess the improvement in computational performance as the number of processing units increased while keeping the problem size constant, demonstrating the efficiency of the km-scale ELM simulation in a large cluster. The second test examined weak scalability, where we increased the workload proportionally with the addition of nodes, using configurations from 1 to 300 nodes. This approach showed the ELM’s ability to maintain performance as the workload expanded. 
\vspace{2mm}
\subsubsection{Strong scaling experiments with a fixed program size}

The strong scaling test is evaluated using the largest experiment (AKSPx10x10x3), which requires at least 6,300 cores (equivalent to 150 Summit nodes), i.e., approximately 3,500 gridcells per core. In this experiment, a fixed number of MPIs are utilized for both the data atmosphere and coupler, corresponding to 3,360 tasks distributed across 80 Summit nodes. This allocation is necessary because the workload associated with data atmosphere (temporal data interpolation) and coupler (data exchange) remains constant once the domain is defined. Additionally, 80 nodes are selected to ensure that the execution time of the atmosphere and coupler does not overshadow that of the land model during performance profiling. Table~\ref{tab5} lists the experiment configuration, and the time spent on model initialization and execution stage (of each component). We measure the performance in two terms: wall time (in second) and simulation-years-per-day (SYPD). 

\begin{table*}[htb]
\caption{Strong scaling experiments with the AKSPx10x10x3 case}
\begin{center}
\begin{tabular}{|c|c|c|c|c|c|}
\hline
Total MPI Task & 6,300 & 12,600 & 25,200 & 50,400 & 100,800 \\
\hline
ELM Configuration &
\begin{tabular}{@{}c@{}}
ATM- 3,360 \\
CPL- 3,360 \\
LND- 6,300  \end{tabular} &
\begin{tabular}{@{}c@{}}
ATM- 3,360 \\
CPL- 3,360 \\
LND- 12,600 \end{tabular} &
\begin{tabular}{@{}c@{}}
ATM- 3,360 \\
CPL- 3,360 \\
LND- 25,200 \end{tabular} &
\begin{tabular}{@{}c@{}}
ATM- 3,360 \\
CPL- 3,360 \\
LND- 50,400 \end{tabular} &
\begin{tabular}{@{}c@{}}
ATM- 3,360 \\
CPL- 3,360 \\
LND- 100,800 \end{tabular} \\
\hline
Initialization & 1,977.47  & 1,591.55  & 1,631.04 & 1,963.68 & 2,060.34 \\ 
\hline
\begin{tabular}{@{}c@{}}
ATM Second \\SYPD \end{tabular} &
\begin{tabular}{@{}c@{}}
132.749 \\
8.92 \end{tabular} &
\begin{tabular}{@{}c@{}}
48.765 \\
24.27 \end{tabular} &
\begin{tabular}{@{}c@{}}
31.120\\
38.03 \end{tabular} &
\begin{tabular}{@{}c@{}}
68.089 \\
17.38 \end{tabular} &
\begin{tabular}{@{}c@{}}
52.85 \\
22.35 \end{tabular} \\
\hline
\begin{tabular}{@{}c@{}}
CPL
Second \\
SYPD \end{tabular} &
\begin{tabular}{@{}c@{}}
142.032\\
8.33 \end{tabular} &
\begin{tabular}{@{}c@{}}
39.008\\
30.34 \end{tabular} &
\begin{tabular}{@{}c@{}}
3.886\\
304.57 \end{tabular} &
\begin{tabular}{@{}c@{}}
1.573\\
752 \end{tabular} &
\begin{tabular}{@{}c@{}}
3.32\\
366.51 \end{tabular} \\
\hline
\begin{tabular}{@{}c@{}}
LND 
Second\\SYPD \end{tabular} &
\begin{tabular}{@{}c@{}}
939.447\\
1.26 \end{tabular} &
\begin{tabular}{@{}c@{}}
388.043\\
3.05 \end{tabular} &
\begin{tabular}{@{}c@{}}
185.725\\
6.37 \end{tabular} &
\begin{tabular}{@{}c@{}}
134.960\\
8.7 \end{tabular} &
\begin{tabular}{@{}c@{}}
102.042\\
11.60 \end{tabular} \\
\hline
\end{tabular}
\label{tab5}
\end{center}
\end{table*}

The model initialization time for the AKSPx10x10x3 cases ranges between 1,600 and 2,100 seconds. The initialization time was affected by the LND initialization. When 6,300 cores (150 nodes) were used, IO operations influenced the performance. While more than 50,400 cores were used, MPI collective communication among all processes became significant. 
The initialization time is around 1,600 seconds when utilizing 12,500 to 25,200 cores, equating to approximately 1,000 to 2,000 gridcells per process. 

During the model execution phase of the data atmosphere model (ATM), a similar situation was observed, prompting an investigation into the I/O performance of these runs. The findings indicate that optimal I/O performance is achieved when utilizing 25,200 cores (or 600 nodes) for I/O, with a maximum of 64 MB allocated for I/O operations through the PnetCDF APIs.

The performance of the coupler (CPL) is enhanced as the number of cores increases. This is because the data exchange between the land and the coupler uses block MPI communications. When 100,800 cores are employed for land, the communication between 3,360 cores of the coupler and 100,800 cores of land experiences a slight increase. However, in general, the execution time of the coupler is not significant when more than 25,200 cores were utilized in the experiments.

\begin{figure}[b]
\centerline{\includegraphics[width=\linewidth]{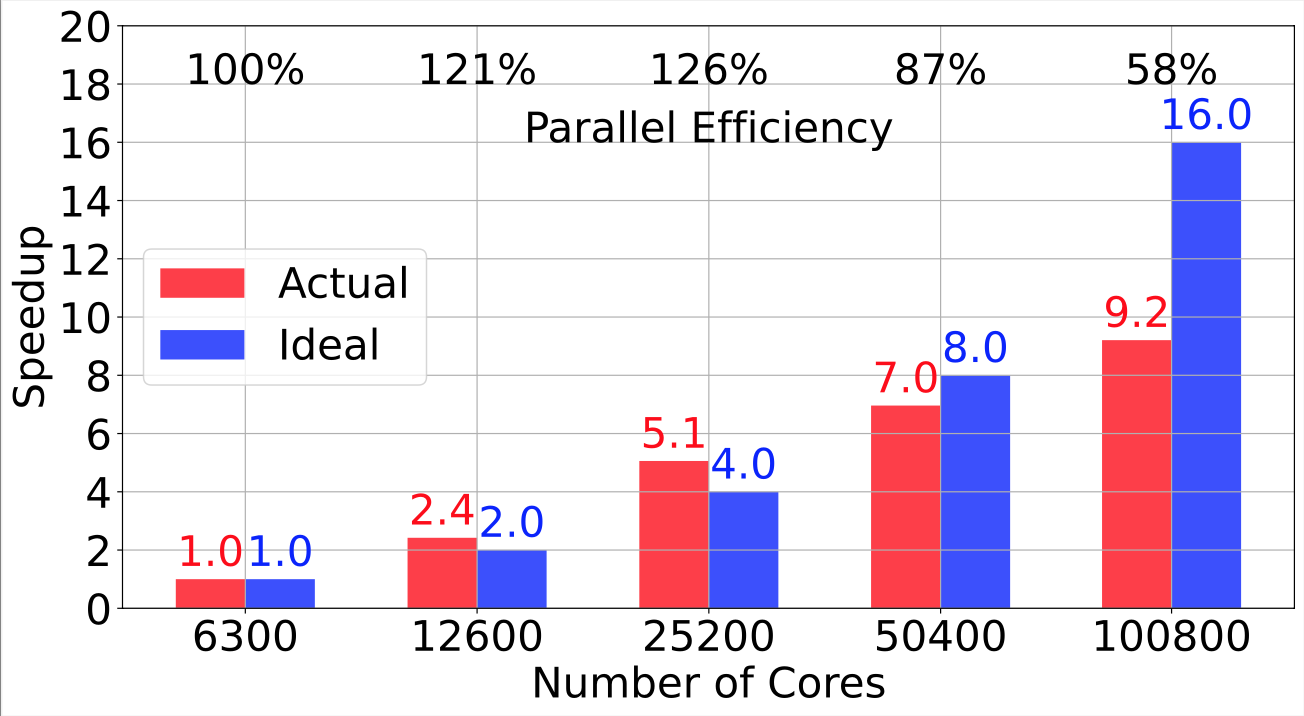}}
\caption{Speedup of the land model in the strong scaling experiments. }
\label{fig_5}
\end{figure}

The land model (LND) demonstrates favorable scalability outcomes, characterized by a significant reduction in execution time from around 940 seconds to 102 seconds when employing 6,00 cores (150 nodes) and 100,800 cores (2,400 nodes), respectively. Fig.~\ref{fig_5} shows the speedup of the land model in the strong scaling experiments, illustrating how LND handles a fixed total workload as the number of cores increases. The x-axis represents the number of cores, ranging from 6,300 to 100,800, while the y-axis indicates the achieved speedup relative to the baseline performance at 6,300 cores, which is set as the reference speedup of 1.0. The blue bars represent the ideal or perfect linear speedup, which scales proportionally with the number of cores. In contrast, the red bars show the actual measured speedup. At the top of the plot, the parallel efficiency—defined as the ratio of the actual speedup to the ideal speedup, expressed as a percentage—is annotated for each tested core count. As is typical in strong scaling scenarios, the parallel efficiency decreases as more resources are applied to the same problem size. Notably, LND maintains an impressive parallel efficiency of 87\% when scaling from 6,300 cores to 50,400 cores. The efficiency has dropped to 58\% when the experiment reaches 100,800 cores where each core is allocated only about 220 gridcells, meaning that the overheads—such as communication costs, synchronization delays, and/or diminishing cache/locality benefits—are reducing the effective gains from additional processing units~\cite{slaughter2020task}.




\vspace{2mm}
\subsubsection{Weak scaling experiments with a fixed workload on individual processes}

\begin{table}[b]
\caption{Weak scaling experiments with the AKSP reference case}
\begin{center}
\begin{tabular}{|c|c|c|c|c|}
\hline
Case &
\makecell{AKSP} &
\makecell{AKSP\\x10} &
\makecell{AKSP\\x10x10} &
\makecell{AKSP\\x10x10x3} \\ \hline

\makecell{ELM \\Configuration} &
\begin{tabular}{@{}c@{}}
ATM-11 \\
CPL-11 \\
LND-42 \end{tabular} &
\begin{tabular}{@{}c@{}}
ATM-112 \\
CPL-112 \\
LND-420 \end{tabular} &
\begin{tabular}{@{}c@{}}
ATM-1,120 \\
CPL-1,120 \\
LND-4,200 \end{tabular} &
\begin{tabular}{@{}c@{}}
ATM-3,360 \\
CPL-3,360 \\
LND-12,600 \end{tabular} \\
\hline
Initialization &
87.056 &
120.692 &
931.272 &
1,591.552 \\
\hline
ATM &
\begin{tabular}{@{}c@{}}
2.395 \\
494.18 \end{tabular}&
\begin{tabular}{@{}c@{}}
11.353 \\
104.25 \end{tabular}&
\begin{tabular}{@{}c@{}}
28.985\\
40.83 \end{tabular}&
\begin{tabular}{@{}c@{}}
48.765\\
24.27 \end{tabular}\\
\hline
CPL &
\begin{tabular}{@{}c@{}}
0.597\\
1,982.52 \end{tabular}&
\begin{tabular}{@{}c@{}}
6.577\\
179.95 \end{tabular}&
\begin{tabular}{@{}c@{}}
8.534\\
138.69 \end{tabular}&
\begin{tabular}{@{}c@{}}
39.008\\
30.34 \end{tabular}\\
\hline
LND &
\begin{tabular}{@{}c@{}}
316.927\\
3.73 \end{tabular}&
\begin{tabular}{@{}c@{}}
374.488\\
3.16 \end{tabular}&
\begin{tabular}{@{}c@{}}
392.896 \\
3.01 \end{tabular}&
\begin{tabular}{@{}c@{}}
388.043 \\
3.05 \end{tabular}\\
\hline
\end{tabular}
\label{tab6}
\end{center}
\end{table}

In the weak scaling testing scenario, an equivalent workload (about 1,700 gridcells) is allocated on each processor/core, down-scaling from AKSP
x10x10x3 on 12,600 cores in Table~\ref{tab5}. Table~\ref{tab6} presents the weak scaling results in terms of wall time (in seconds)
and SYPD, using a baseline configuration AKSP on 42 cores (1 node) and then scaling up to AKSPx10, AKSPx10x10, and AKSPx10x10x3 on 420, 4,200, and 12,600 cores, respectively. The number of cores utilized by ATM and CPL is the same. The ratio of cores used in ATM/CPL to those in LND closely mirrors the ASKPx10x10x3 scenario, that is 3,360/12,600.

It is worth explaining that the model initialization process exhibited poor scalability. The time required for initialization increased significantly, from 87 seconds in the AKSP case to approximately 1,600 seconds in the AKSPx10x10x3 case, representing an increase of approximately 18.4 times. One primary factor that contributed to this performance degradation is the land model initialization. When 12,600 cores were used (with large input dataset) in the largest case, the IO operations and MPI collective communication influenced the scalability significantly. It is one issue we will address in future work.


\vspace{2mm}
During the model execution phase, the data atmosphere model (ATM)'s execution time increased significantly from 2.4 seconds in case AKSP to 48.76 seconds in case AKSPx10x10x3, representing a 20.3-fold increase. The increase is primarily attributed to a much larger input dataset in the AKSPx10x10x3 case. The default I/O configuration involves the utilization of 10 to 100 nodes for I/O read operations. The coupler (CPL)'s execution time surged from 0.6 to 39.0 seconds, a 65.0-fold increase, predominantly due to synchronized data communications between the data atmosphere, coupler, and land model.

\vspace{1mm}
The land model demonstrates superior scalability. In the AKSP scenario, with 42 MPI tasks on a single Summit node, the land model required approximately 317.0 seconds. The land model execution time in the AKSPx10 case increased to approximately 375 seconds, mainly due to differences in data synchronization and aggregation among tasks on single-node (AKSP case) and multiple-node at each ELM simulation timestep. Notably, in larger cases such as AKSPx10, AKSPx10x10, and AKSPx10x10x3, the land execution time remains consistent at around 380 seconds. These larger cases exhibit commendable weak scalability performance, with similar execution times despite growing problem sizes. This outcome is unsurprising, as the current land model is a gridcell-independent application with workload-balanced partitioning and no lateral interactions among gridcells.

\begin{figure}[b]
\centerline{\includegraphics[width=\linewidth]{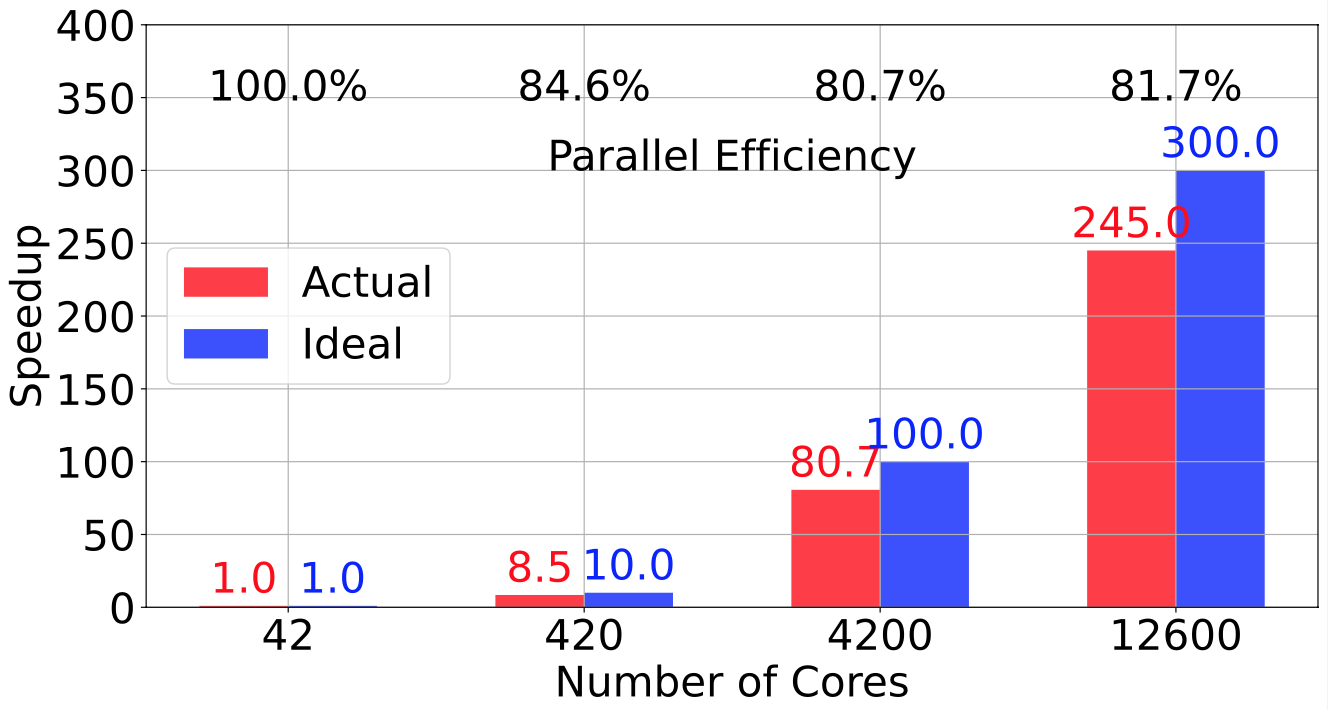}}
\caption{Scaled speedup of the land model in the weak scaling experiments.}
\label{fig_6}
\end{figure}

\vspace{1mm}
Figure~\ref{fig_6} illustrates the speedup achieved in the weak scaling experiments. The x-axis displays the number of cores, beginning with the baseline configuration of 42 cores (1 node) and scaling up to 12,600 cores. The y-axis indicates the resulting speedup. Similarly to Figure~\ref{fig_5}, the red bars represent the ideal speed-up, the blue bars show the actual speedup, and the parallel efficiency is labeled at each scaling point. The results demonstrate that LND achieves high efficiencies, exceeding 80\% in all examined configurations. In other words, as the problem size and the number of cores increase, LND continues to run efficiently without a drastic drop in per-core performance. The slight performance degradation observed may be attributed to the current suboptimal implementation of large grid partitioning and surface data structure initialization.


In the study, another considerable issue identified was the scalability of I/O, specifically concerning write operations. Performance of the write (both history and restart files), measured through the SCORPIO APIs, was recorded at the end of the simulation. In the AKSP case, 15.14 GB was generated in 21.51 seconds, resulting in an average write bandwidth of 671.8 MB/s. The write bandwidth for the AKSPx10 and AKSPx10x10 cases reached 3.33 GB/s and 13.33 GB/s respectively. However, the AKSPx10x10x3 case required using a large I/O buffer (64MB) on each process to sustain execution. This case generated 4540.54 GB in 503.12 seconds, achieving an average write bandwidth of 8.4GB/s.
\vspace{1mm}
\section{Conclusions and future work}

The km-scale ELM is a complex application with over half a million lines of code, a thousand subroutines, and a few thousands of global multidimensional arrays. The successful execution of the strong and weak scaling experiments represents a significant advancement in km-scale ELM simulation, enabling researchers to capture the complexity and variability of terrestrial ecosystem processes over North America at kilometer spatial resolutions.   

This study presents the first full-capacity ELM simulations on a continental scale, adopting a spatial resolution of 1 km × 1 km using a maximum of 100,800 CPU cores. It is worth mentioning that the current km-scale ELM also compasses computational competence for global simulation at a resolution of 3 km × 3 km.   

The km-scale ELM is fully compatible with the CIME framework and has demonstrated substantial improvements in
performance and scalability, addressing key challenges associated with data management and processing in high-resolution modeling. Moreover, the study presented a robust approach to support ELM simulations for future integration with fully-coupled E3SM simulation and can conveniently be reconfigured for km-scale ELM simulation over area of interest, ranging from a single land gridcell to the entire domain of North America. 

\vspace{1mm}
Moving forward, efforts from a computing perspective will focus on resolving identified computational issues. This includes enhancing model initialization performance, developing new domain partition schemes, and optimizing data structures and functions to reduce memory usage and improve computing efficiency.


\vspace{1mm}
The development of high-fidelity surface properties datasets at kilometer scale is a critical requirement. Future endeavors will also involve upscaling kilometer-scale ELM simulations utilizing both CPU and GPU resources. Additionally, advanced data analysis techniques will be developed to analyze outputs of kilometer-scale ELM efficiently. Transient simulations will be conducted to reveal the ecosystem states in the present time, then ELM land diagnostic analysis and observation data can be used to verify the model outcomes. 

\bibliographystyle{IEEEtrans}
\bibliography{myref,io}

\begin{thebibliography}{10}
\providecommand{\url}[1]{#1}
\csname url@samestyle\endcsname
\providecommand{\newblock}{\relax}
\providecommand{\bibinfo}[2]{#2}
\providecommand{\BIBentrySTDinterwordspacing}{\spaceskip=0pt\relax}
\providecommand{\BIBentryALTinterwordstretchfactor}{4}
\providecommand{\BIBentryALTinterwordspacing}{\spaceskip=\fontdimen2\font plus
\BIBentryALTinterwordstretchfactor\fontdimen3\font minus \fontdimen4\font\relax}
\providecommand{\BIBforeignlanguage}[2]{{%
\expandafter\ifx\csname l@#1\endcsname\relax
\typeout{** WARNING: IEEEtran.bst: No hyphenation pattern has been}%
\typeout{** loaded for the language `#1'. Using the pattern for}%
\typeout{** the default language instead.}%
\else
\language=\csname l@#1\endcsname
\fi
#2}}
\providecommand{\BIBdecl}{\relax}
\BIBdecl

\bibitem{golaz2019doe}
J.-C. Golaz, P.~M. Caldwell, L.~P. Van~Roekel, M.~R. Petersen, Q.~Tang, J.~D. Wolfe, G.~Abeshu, V.~Anantharaj, X.~S. Asay-Davis, D.~C. Bader \emph{et~al.}, ``The doe e3sm coupled model version 1: Overview and evaluation at standard resolution,'' \emph{Journal of Advances in Modeling Earth Systems}, vol.~11, no.~7, pp. 2089--2129, 2019.

\bibitem{caldwell2019}
\BIBentryALTinterwordspacing
P.~M. Caldwell, A.~Mametjanov, Q.~Tang, L.~P. Van~Roekel, J.-C. Golaz, W.~Lin, D.~C. Bader, N.~D. Keen, Y.~Feng, R.~Jacob, M.~E. Maltrud, A.~F. Roberts, M.~A. Taylor, M.~Veneziani, H.~Wang, J.~D. Wolfe, K.~Balaguru, P.~Cameron-Smith, L.~Dong, S.~A. Klein, L.~R. Leung, H.-Y. Li, Q.~Li, X.~Liu, R.~B. Neale, M.~Pinheiro, Y.~Qian, P.~A. Ullrich, S.~Xie, Y.~Yang, Y.~Zhang, K.~Zhang, and T.~Zhou, ``The doe e3sm coupled model version 1: Description and results at high resolution,'' \emph{Journal of Advances in Modeling Earth Systems}, vol.~11, no.~12, pp. 4095--4146, 2019. [Online]. Available: \url{https://agupubs.onlinelibrary.wiley.com/doi/abs/10.1029/2019MS001870}
\BIBentrySTDinterwordspacing

\bibitem{burrows2020doe}
S.~Burrows, M.~Maltrud, X.~Yang, Q.~Zhu, N.~Jeffery, X.~Shi, D.~Ricciuto, S.~Wang, G.~Bisht, J.~Tang \emph{et~al.}, ``The doe e3sm v1. 1 biogeochemistry configuration: Description and simulated ecosystem-climate responses to historical changes in forcing,'' \emph{Journal of Advances in Modeling Earth Systems}, vol.~12, no.~9, p. e2019MS001766, 2020.

\bibitem{soares2024high}
P.~M. Soares, F.~Johannsen, D.~C. Lima, G.~Lemos, V.~A. Bento, and A.~Bushenkova, ``High-resolution downscaling of cmip6 earth system and global climate models using deep learning for iberia,'' \emph{Geoscientific Model Development}, vol.~17, no.~1, pp. 229--259, 2024.

\bibitem{Donahue2024}
\BIBentryALTinterwordspacing
A.~S. Donahue, P.~M. Caldwell, L.~Bertagna, H.~Beydoun, P.~A. Bogenschutz, A.~M. Bradley, T.~C. Clevenger, J.~Foucar, C.~Golaz, O.~Guba, W.~Hannah, B.~R. Hillman, J.~N. Johnson, N.~Keen, W.~Lin, B.~Singh, S.~Sreepathi, M.~A. Taylor, J.~Tian, C.~R. Terai, P.~A. Ullrich, X.~Yuan, and Y.~Zhang, ``To exascale and beyond—the simple cloud-resolving e3sm atmosphere model (scream), a performance portable global atmosphere model for cloud-resolving scales,'' \emph{Journal of Advances in Modeling Earth Systems}, vol.~16, no.~7, p. e2024MS004314, 2024, e2024MS004314 2024MS004314. [Online]. Available: \url{https://agupubs.onlinelibrary.wiley.com/doi/abs/10.1029/2024MS004314}
\BIBentrySTDinterwordspacing

\bibitem{omega}
``Omega- future e3sm ocean model,'' \url{https://climatemodeling.science.energy.gov/news/omega-future-e3sm-ocean-model}, accessed: 2010-09-30.

\bibitem{xu2017web}
Y.~Xu, D.~Wang, T.~Janjusic, W.~Wu, Y.~Pei, and Z.~Yao, ``A web-based visual analytic framework for understanding large-scale environmental models: A use case for the community land model,'' \emph{Procedia Computer Science}, vol. 108, pp. 1731--1740, 2017.

\bibitem{zheng2019xscan}
W.~Zheng, D.~Wang, and F.~Song, ``Xscan: an integrated tool for understanding open source community-based scientific code,'' in \emph{International Conference on Computational Science}.\hskip 1em plus 0.5em minus 0.4em\relax Springer, 2019, pp. 226--237.

\bibitem{wang2022towards}
D.~Wang, P.~Schwartz, F.~Yuan, P.~Thornton, and W.~Zheng, ``Towards ultra-high-resolution e3sm land modeling on exascale computers,'' \emph{Computing in Science \& Engineering}, no.~01, pp. 1--14, 2022.

\bibitem{YUAN2023102145}
F.~Yuan, D.~Wang, S.-C. Kao, M.~Thornton, D.~Ricciuto, V.~Salmon, C.~Iversen, P.~Schwartz, and P.~Thornton, ``An ultrahigh-resolution e3sm land model simulation framework and its first application to the seward peninsula in alaska,'' \emph{Journal of Computational Science}, vol.~73, p. 102145, 2023.

\bibitem{schwartz2022spel}
P.~Schwartz, D.~Wang, F.~Yuan, and P.~Thornton, ``Spel: Software tool for porting e3sm land model with openacc in a function unit test framework,'' in \emph{Accelerator Programming--WACCPD 2022: 9th Workshop on Accelerator Programming Using Directives, Dallas, USA, Nov 18, 2022, Proceedings}.\hskip 1em plus 0.5em minus 0.4em\relax Springer, 2022, pp. 1--14.

\bibitem{schwartz2022developing}
P.~Schwartz, D.~Wang, and P.~Thornton, ``Developing an elm ecosystem dynamics model on gpu with openacc,'' in \emph{22nd International Conference on Computational Science--ICCS 2022, London, UK, June 21--23, 2022, Proceedings, Part II}.\hskip 1em plus 0.5em minus 0.4em\relax Springer, 2022, pp. 291--303.

\bibitem{hao2022impacts}
D.~Hao, G.~Bisht, M.~Huang, P.-L. Ma, T.~Tesfa, W.-L. Lee, Y.~Gu, and L.~R. Leung, ``Impacts of sub-grid topographic representations on surface energy balance and boundary conditions in the e3sm land model: A case study in sierra nevada,'' \emph{Journal of Advances in Modeling Earth Systems}, vol.~14, no.~4, p. e2021MS002862, 2022.

\bibitem{thornton2014daymet}
P.~E. Thornton, M.~M. Thornton, B.~W. Mayer, N.~Wilhelmi, Y.~Wei, R.~Devarakonda, and R.~B. Cook, ``Daymet: Daily surface weather data on a 1-km grid for north america, version 2.'' Oak Ridge National Lab.(ORNL), Oak Ridge, TN (United States), Tech. Rep., 2014.

\bibitem{thornton2021gridded}
P.~E. Thornton, R.~Shrestha, M.~Thornton, S.-C. Kao, Y.~Wei, and B.~E. Wilson, ``Gridded daily weather data for north america with comprehensive uncertainty quantification,'' \emph{Scientific Data}, vol.~8, no.~1, pp. 1--17, 2021.

\bibitem{han2023global}
Q.~Han, Y.~Zeng, L.~Zhang, C.~Wang, E.~Prikaziuk, Z.~Niu, and B.~Su, ``Global long term daily 1 km surface soil moisture dataset with physics informed machine learning,'' \emph{Scientific Data}, vol.~10, no.~1, p. 101, 2023.

\bibitem{Childs2012visit}
\BIBentryALTinterwordspacing
H.~Childs, E.~Brugger, B.~Whitlock, J.~Meredith, S.~Ahern, D.~Pugmire, K.~Biagas, M.~C. Miller, C.~Harrison, G.~H. Weber, H.~Krishnan, T.~Fogal, A.~Sanderson, C.~Garth, E.~W. Bethel, D.~Camp, O.~Rubel, M.~Durant, J.~M. Favre, and P.~Navratil, ``{High Performance Visualization--Enabling Extreme-Scale Scientific Insight},'' 2012. [Online]. Available: \url{https://visit.llnl.gov}
\BIBentrySTDinterwordspacing

\bibitem{QGIS_software}
\BIBentryALTinterwordspacing
{QGIS Development Team}, \emph{QGIS Geographic Information System}, QGIS Association, 
\BIBentrySTDinterwordspacing

\bibitem{scropio2020}
J.~Krishna and D.~Wu, ``Software for caching output and reads for parallel i/o,'' \url{https://github.com/E3SM-Project/scorpio}, 2020.

\bibitem{Hartnett2021PIO}
E.~Hartnett and J.~Edwards, ``The parallelio (pio) c/fortran libraries for scalable hpc performance,'' 01 2021.

\bibitem{sun:2007:lustre}
SUN, ``{High-Performance Storage Architecture and Scalable Cluster File System},'' Sun Microsystems, Inc, Tech. Rep., 2007.

\bibitem{schmuck:2002:gpfs}
F.~Schmuck and R.~Haskin, ``{GPFS: A Shared-Disk File System for Large Computing Clusters},'' in \emph{Proceedings of the 1st USENIX Conference on File and Storage Technologies}, ser. FAST’02.\hskip 1em plus 0.5em minus 0.4em\relax USA: USENIX Association, 2002, pp. 231--244.

\bibitem{wang2023data}
D.~Wang, F.~Yuan, P.~E. Thornton, P.~Schwartz, S.-C. Kao, M.~Thornton, and D.~M. Ricciuto, ``A data toolkit for ultrahigh-resolution e3sm simulation on massively parallel processing systems,'' Oak Ridge National Laboratory (ORNL), Oak Ridge, TN (United States), Tech. Rep., 2023.

\bibitem{kao2024gridded}
S.-C. Kao, M.~Thornton, P.~E. Thornton, and R.~Shrestha, ``Gridded sub-daily climate forcings for north america based on daymet and gswp3 (daymet-gswp3),'' Oak Ridge National Laboratory (ORNL), Oak Ridge, TN (United States). Oak~…, Tech. Rep., 2024.

\bibitem{ngee-arctic}
``Next-generation ecosystem experiments – arctic,'' \url{https://https://ess.science.energy.gov/ngee-arctic/}, accessed: 2010-10-30.

\bibitem{slaughter2020task}
E.~Slaughter, W.~Wu, Y.~Fu, L.~Brandenburg, N.~Garcia, W.~Kautz, E.~Marx, K.~S. Morris, Q.~Cao, G.~Bosilca \emph{et~al.}, ``Task bench: A parameterized benchmark for evaluating parallel runtime performance,'' in \emph{SC20: International Conference for High Performance Computing, Networking, Storage and Analysis}.\hskip 1em plus 0.5em minus 0.4em\relax IEEE, 2020, pp. 1--15.

\end{thebibliography}
\end{document}